\documentclass[]{aa}
\usepackage{psfig}
\begin{document}
\title{Oxygen and Magnesium Abundance in the Ultra-Metal-Poor Giants CS\,22949-037 and CS\,29498-043: 
Challenges in Models of Atmospheres.}
\author{G.~Israelian\inst{1}
\and  N.~Shchukina\inst{2}
\and  R.~Rebolo\inst{1,3}
\and  G.~Basri\inst{4}
\and  J. I.~Gonz\'alez Hern\'andez\inst{1}
\and  T.~Kajino\inst{5}
}
\offprints{G.Israelian (gil@iac.es)}
\institute{Instituto de Astrof{\'\i}sica de Canarias,, E-38205 La Laguna,
Tenerife, Spain
\and
Main Astronomical Observatory, National Academy of Sciences, 
03680~Kyiv-127, Ukraine
\and
Consejo Superior de Investigaciones Cient\'\i fcas, Spain
\and
Astronomy Department, University of California, Berkeley, California 94720, USA
\and
National Astronomical Observatory of Japan, 2-21-1, Osawa, Mitaka, Tokyo 181-8588, Japan
}
\date{Received; accepted}

\titlerunning{Oxygen and Magnesium in CS\,22949-037 and CS\,29498-043}

\abstract{
We report the results of a non-LTE Fe, O and Mg abundance analysis of the carbon-nitrogen-rich 
ultra-metal-poor giants CS\,29498--043 and CS\,22949--037. The abundance of oxygen has 
been derived from measurements of the oxygen triplet at 7771--5 \AA\ in high resolution  
spectra obtained with KeckI/HIRES and the forbidden line [O\,{\sc i}] 6300 \AA\ detected in 
the TNG/SARG spectra of CS\,29498-043. Detailed non-LTE analysis of Fe lines has provided reliable
stellar parameters which, however, do not resolve the oxygen abundance conflict as derived
from the O\,{\sc i} 7771-5 \AA\ triplet and the [O\,{\sc i}] 6300 \AA\ forbidden lines.
We obtained the following oxygen abundance: for CS\,22949--037 [O/Fe] =3.13, 1.95; and
for CS\,29498--043; [O/Fe]=3.02, 2.49, based on the O\,{\sc i} 7771--5 \AA\ triplet and
the [O\,{\sc i}] 6300 \AA\ forbidden line, respectively.  
A similar conflict appears to exist between the forbidden resonance line Mg\,{\sc i} 4571 \AA\ 
and several subordinate lines, such as Mg\,{\sc i} 5172 and 5183 \AA. 
Our analysis demonstrates the failure of standard plane--parallel atmosphere models to 
describe the physical conditions in the line-forming regions of these ultra-metal-poor giants.

}

\maketitle

\section{Introduction}

Surveys of metal-poor stars (Beers et al. 1992) and their abundance studies (Ryan, Norris \& Beers 1996; 
Norris, Ryan \& Beers\,1999) are aimed at investigating the chemical evolution of the Galaxy and 
the nucleosynthetic yields of supernovae.
Despite the numerous studies in this field, the current situation with abundance 
trends of various $\alpha$-elements is very confusing. Observations of Stephens \& Boesgaard (2002)
demonstrate that the [$\alpha$/Fe] ratios for  Ca, Si, Ti and Mg do not show a flat {\itshape
plateau} at [Fe/H] $< -1$, as was claimed in previous studies. The results obtained by 
Idiart \& Thevenin\,(1999) and Stephens \& Boesgaard (2002) cannot be called  {\it consistent} 
with the analysis presented by Carretta et al.\,(2002) and many others (see McWilliam 1997). 
The situation with oxygen is far from being resolved (Israelian, Garc\'\i a L\'opez \& Rebolo 
\,1998; Israelian et al.\,2001; Nissen et al.\ 2002; Takeda 2003; 
Fulbright \& Johnson\ 2003), while a new debate over the 
sulfur abundance in metal-poor stars has already emerged (Israelian \& Rebolo 2001; 
Takada-Hidai et al.\ 2002; Nissen et al.\ 2003). These and many other studies clearly 
show  that  [$\alpha$/Fe] $>$ 0 for the great majority of  metal-poor stars in the Galaxy. 
However, it is hard to speak of any trend when the abundance ratios in many stars computed by 
different authors disagree by more than 0.3--0.4 dex. There are many aspects to this serious 
problem, and we shall not discuss them further here. Abundance analysis of ultra-metal-poor stars with 
[Fe/H] $< -$3 gives rise to even more enigmas into this field. 

It is well known that the chemical composition of the atmospheres of halo dwarfs is not 
altered by any internal mixing and therefore provides a good opportunity to constrain 
Galactic chemical evolution models. Unfortunately, most of the known ultra-metal-poor stars are 
not dwarfs but giants (Beers et al. 1992), which pose two serious problems. First, the atmospheric 
parameters of giants are more uncertain and second, their surfaces can be polluted by enriched 
material that has been either dredged from the stellar interior or transferred from a companion star. 
Oxygen is a key element in this scheme as it can help to distinguish between pristine and
pollution origins of other elements and also show which of the
aforementioned processes was dominant. There have been intensive investigations of the oxygen 
abundances in halo stars over the last five years. Abundances derived from near-UV OH lines 
in metal-poor dwarf stars (Israelian et al.\,1998; Boesgaard et al. 1999; 
Israelian et al.\,2001) show that the [O/Fe] ([O/Fe] = log(O/Fe)$_\star$--log (O/Fe)$_\odot$)
ratio increases from 0 to 1 between [Fe/H] = 0 and $-3$. 
The abundances derived from low-excitation OH lines agreed well with those derived 
from high-excitation lines of the O\,{\sc i} triplet at 7771--5 \AA\ (Israelian et al. 1998, 2001; 
Boesgaard et al. 1999 ; Nissen et al.\,2002). It seems that even the [O\,{\sc i}] forbidden line 
at 6300 \AA\ supports the ``quasi-linear'' trend of [O/Fe] (Nissen et al.\,2002) when standard
1D atmospheric models are employed. While some authors claim a good agreement between the 
forbidden line [O\,{\sc i}] 6300 \AA\ and the near-IR triplet (Mishenina et al. 2000; 
Nissen et al.\,2002), others suggest the opposite (Carretta, Gratton, \& Sneden 2000). 
In a recent study Takeda (2003) found that the disagreement between the triplet and the
forbidden line tends to be larger for cool giants. Fulbright \& Johnson (2003) support this
conclusion in a detailed study of 55 subgiants and giants. These authors conclude that it is 
impossible to resolve the disagreement in the two indicators without adopting an ad hoc
temperature scale that is incompatible with standard temperature scales such as IRFM and H$\alpha$.
In fact, the [O/Fe] trend obtained based on the ad hoc scale (see Fig.\ 13 of Fulbright \& Johnson
2003) is identical to the trends presented by Israelian et al. (2001) and Nissen et al. (2002). 
There is no {plateau} at [O/Fe] = 0.5. 
Despite considerable observational effort the trend of the [O/Fe] ratio in the halo is 
still unclear. However, the latest studies (Israelian et al. 2001; Nissen et al. 2002; Takeda 2003; 
Fulbright \& Johnson 2003) suggest that dwarfs provide more reliable and consistent abundances 
than giants. It is not clear how the 3D effects will resolve this conflict since the latter
do not predict an agreement between different oxygen abundance indicators in dwarfs 
(Asplund \& Garc\'\i a P\'erez 2002).

McWilliam et al.\,(1995) were the first to carry out a detailed spectroscopic analysis of 
CS\,22949-037 and to confirm that the star is very metal-poor with an $\alpha$-element excess. 
Furthermore,  Depagne et al.\ (2002) performed a more detailed investigation of this
object and found a large excess of oxygen ([O/Fe] = 2.0) and 
sodium ([Na/Fe] = 2.1). Zero-heavy-element supernovae models with fall back 
have been invoked in order to interpret the elemental abundance ratios in this star.
Aoki et al.\, (2002) have presented a detailed analysis of another 
ultra-metal-poor giant CS\,29498--043 with a very high abundance excess of [Mg/Fe] = 1.81.  
Both,  CS\,22949--037 and CS\,29498--043 exhibit a large overabundance of N and C but show no 
significant enhancement of neutron-capture elements. 
It is possible that the surfaces of these stars have been polluted by enriched material, 
either dredged from the star's inner core or transferred from a companion star. 
The abundances of $\alpha$-elements could be used to discriminate in favour of one 
of these hypotheses or to confirm a pristine origin. The detailed comparison of elemental 
abundances may provide important constraints on the properties of the first supernova progenitors.

In this article we present observations of the oxygen triplet in CS\,22949--037 and CS\,29498-043, as
well as the detection of the  forbidden line [O\,{\sc i}] 6300 \AA\ in CS\,29498--043. 
The stellar parameters and the abundances of iron, oxygen and magnesium were derived in non-LTE. 
We report a significant discrepancy between the abundances derived from the
oxygen triplet and the forbidden line. The conflict cannot be resolved under any circumstances, at least for
CS\,22949-037. A similar conflict was found for Mg. We question the validity of standard plane--parallel 
models of atmospheres employed in the present analysis.

\section{Observations}

The observations of CS\,29498--043 and CS\,22949--037 were performed on 2002 October 31 and 
2002 May 20 (only CS\,22949--037) at the Keck\,I using the high-resolution spectrograph HIRES and 
the TEK 2048 $\times$ 2048 pixel$^{2}$ CCD.  A 1.1 arcsec entrance slit provided a resolving power 
$R \sim$60\,000. A red wavelength setting was used to observe oxygen triplet lines at 7771--5 \AA.
The average signal-to-noise (S/N) ratio of the combined spectrum near 7770 \AA\ was S/N
$>$ 100. Five spectra of CS\,29498--043 with a total exposure
time 13\,500 s were obtained on 2002 September 26 and 27 with the configuration 
CD4 at TNG/SARG (La Palma). A resolving power of 29\,000 was achieved with a 1.6 arcsec slit.
The [O\,{\sc i}] 6300 \AA\ was not blended with any telluric features in this spectral 
window and the sky emission at 6300 \AA\ was carefully subtracted using the off-slit spectra. 
The forbidden line with an equivalent width EW  = 60 $\pm$ 10 m\AA\ was observed in every exposure, providing 
an independent confirmation of the detection. Rotational velocity of CS\,29498--043 derived from 
the triplet (HIRES) and the forbidden line (SARG) was 10 $\pm$ 3 and 8 $\pm$ 3 km\,s$^{-1}$, respectively. 
The S/N ratio 30 was reached near 6300 \AA\ in the unbind spectrum of CS\,29498--043. However, 
given the rotational velocity of the star we could  apply a factor of two bining 
providing S/N $\sim$ 40 to the final spectrum.  All the spectra were 
reduced using standard {\sc iraf}\footnote{{\sc iraf} is distributed by the National Optical 
Astronomical Observatories, which is operated by the Association of Universities for 
Research in Astronomy, Inc., under contract with the National Science Foundation, USA.} 
procedures (bias subtraction, flat-field correction, and extraction of one-dimensional spectra).
Different spectra for each object were co-added before wavelength calibration
and continuum normalization.

\begin{figure*}
\psfig{figure=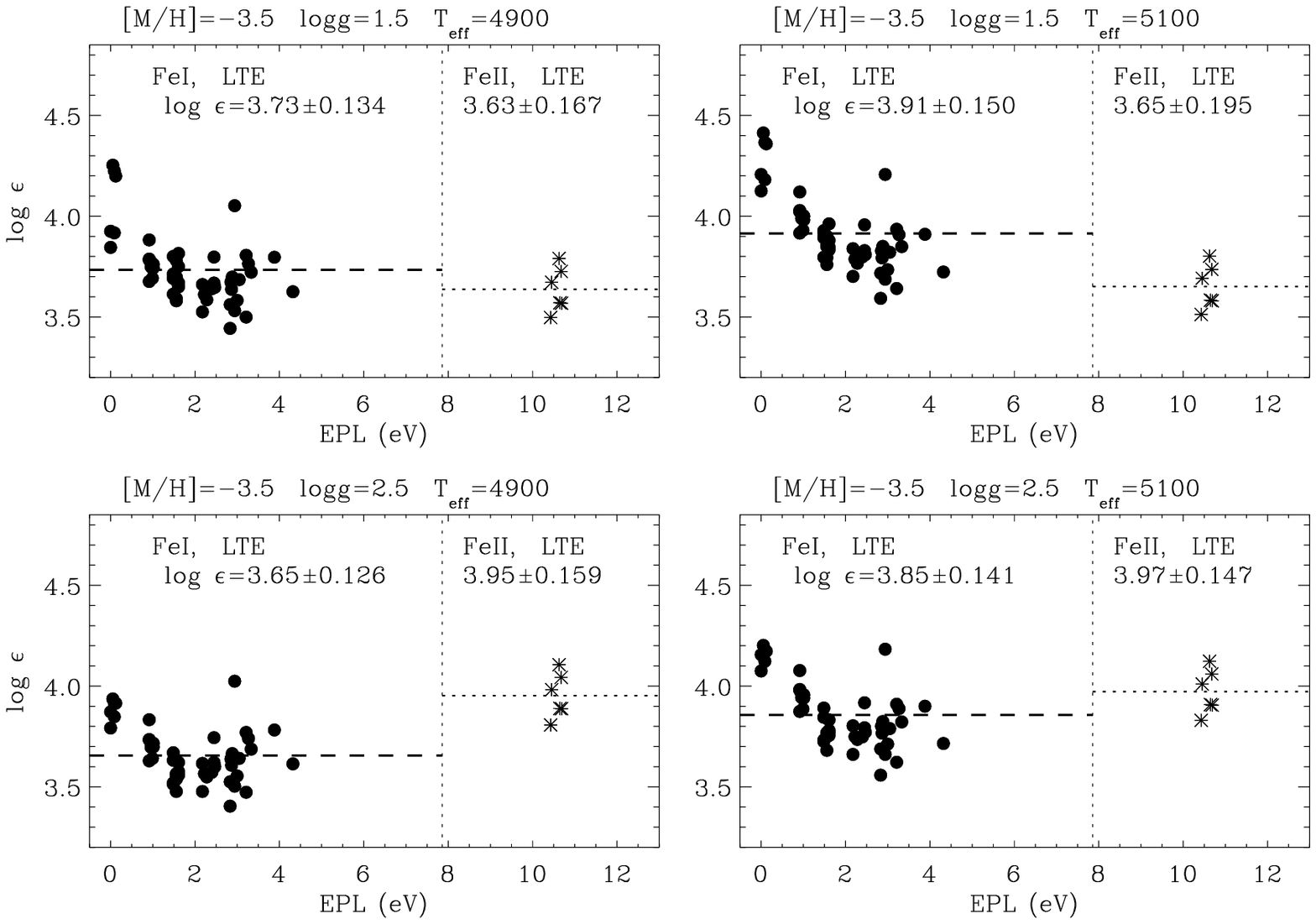,width=11.0cm,height=9.5cm,angle=360}
\caption[]{Results of LTE Fe abundance determination for the Fe\,{\sc i} and Fe\,{\sc ii} 
lines in CS\,22949--037 for a small grid of model atmospheres.}
\end{figure*}

\begin{figure*}
\psfig{figure=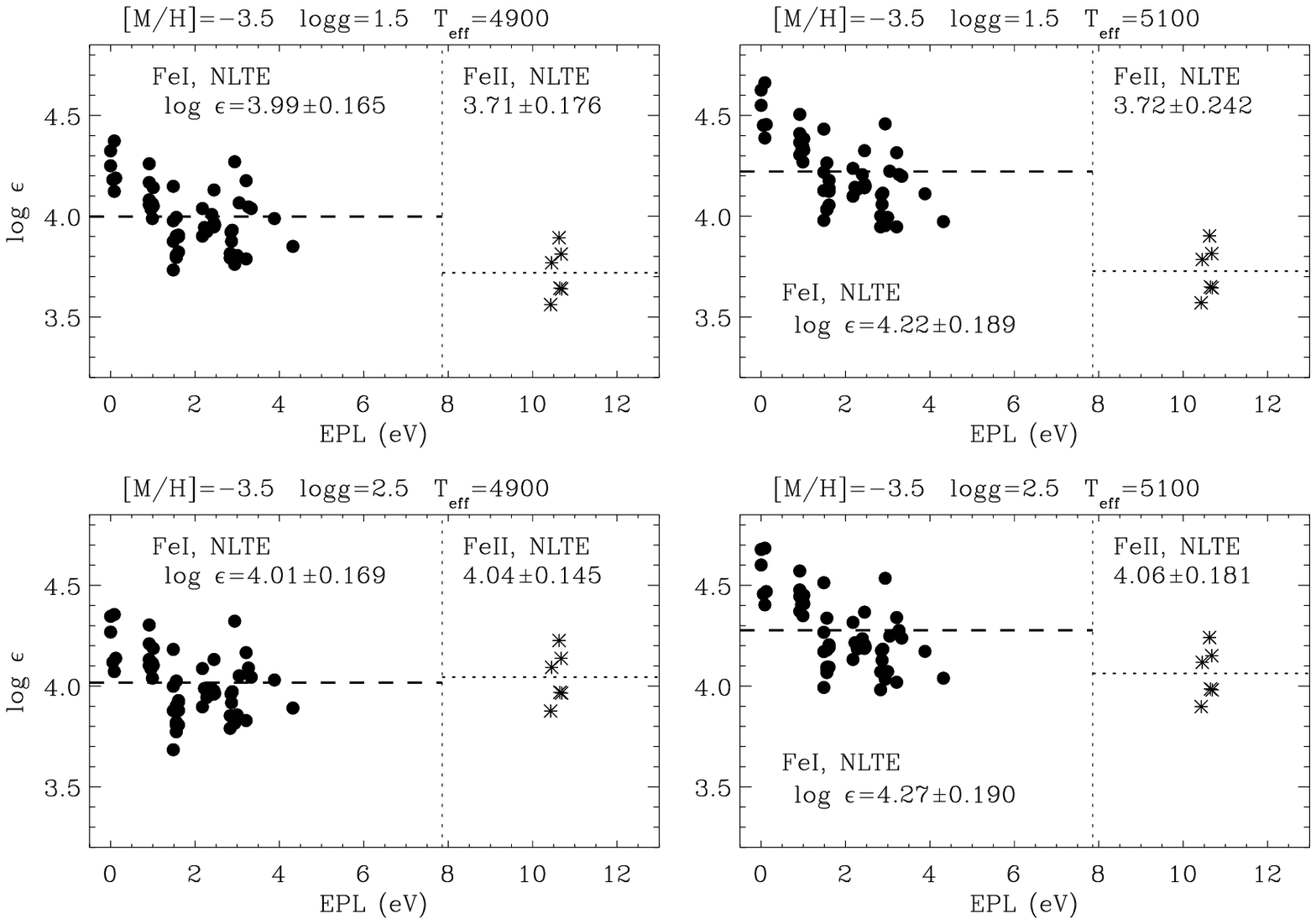,width=11.0cm,height=9.5cm,angle=360}
\caption[]{The same as in Fig. 1 but for a non-LTE Fe analysis.}
\end{figure*}

\begin{figure*}
\psfig{figure=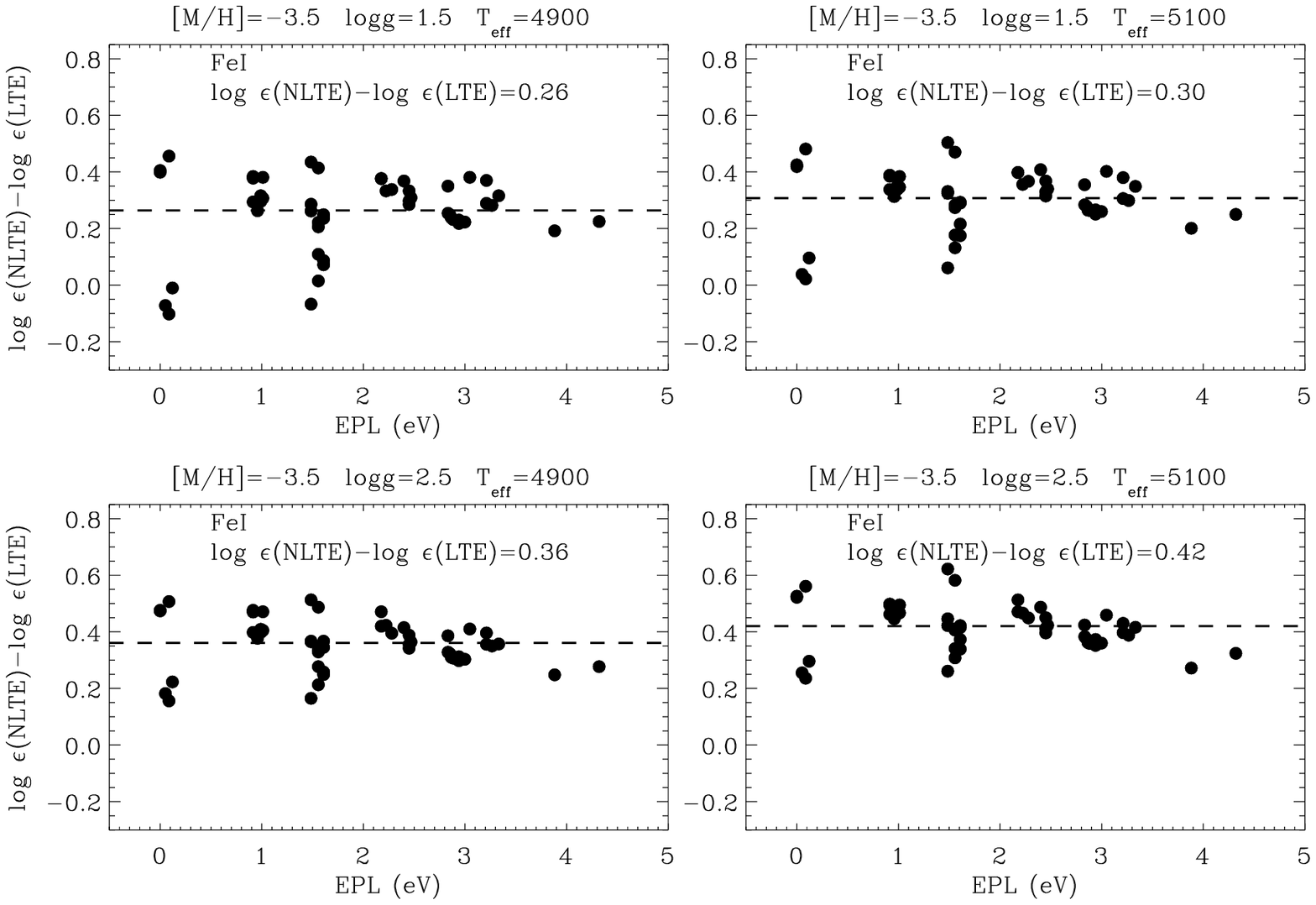,width=11.0cm,height=9.5cm,angle=360}
\caption[]{Difference between non-LTE and LTE abundances as a function 
of $\chi$ in CS\,22949--037.}
\end{figure*}

\begin{figure*}
\psfig{figure=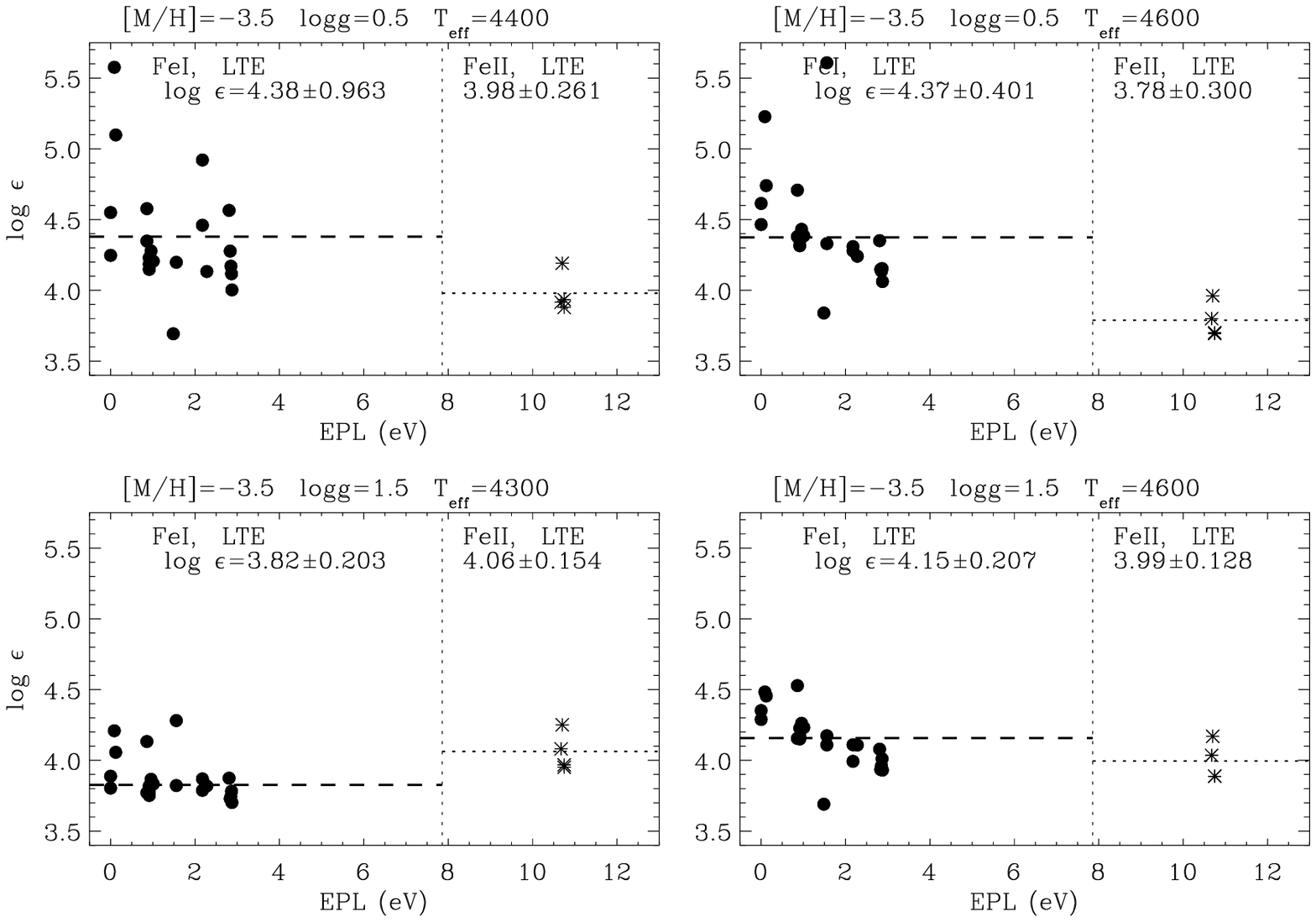,width=11cm,height=9.5cm,angle=360}
\caption[]{Results of LTE Fe abundance determination for Fe\,{\sc i} and Fe\,{\sc ii} 
lines in CS\,29498--043 
for a small grid of model atmospheres.}
\end{figure*}

\begin{figure*}
\psfig{figure=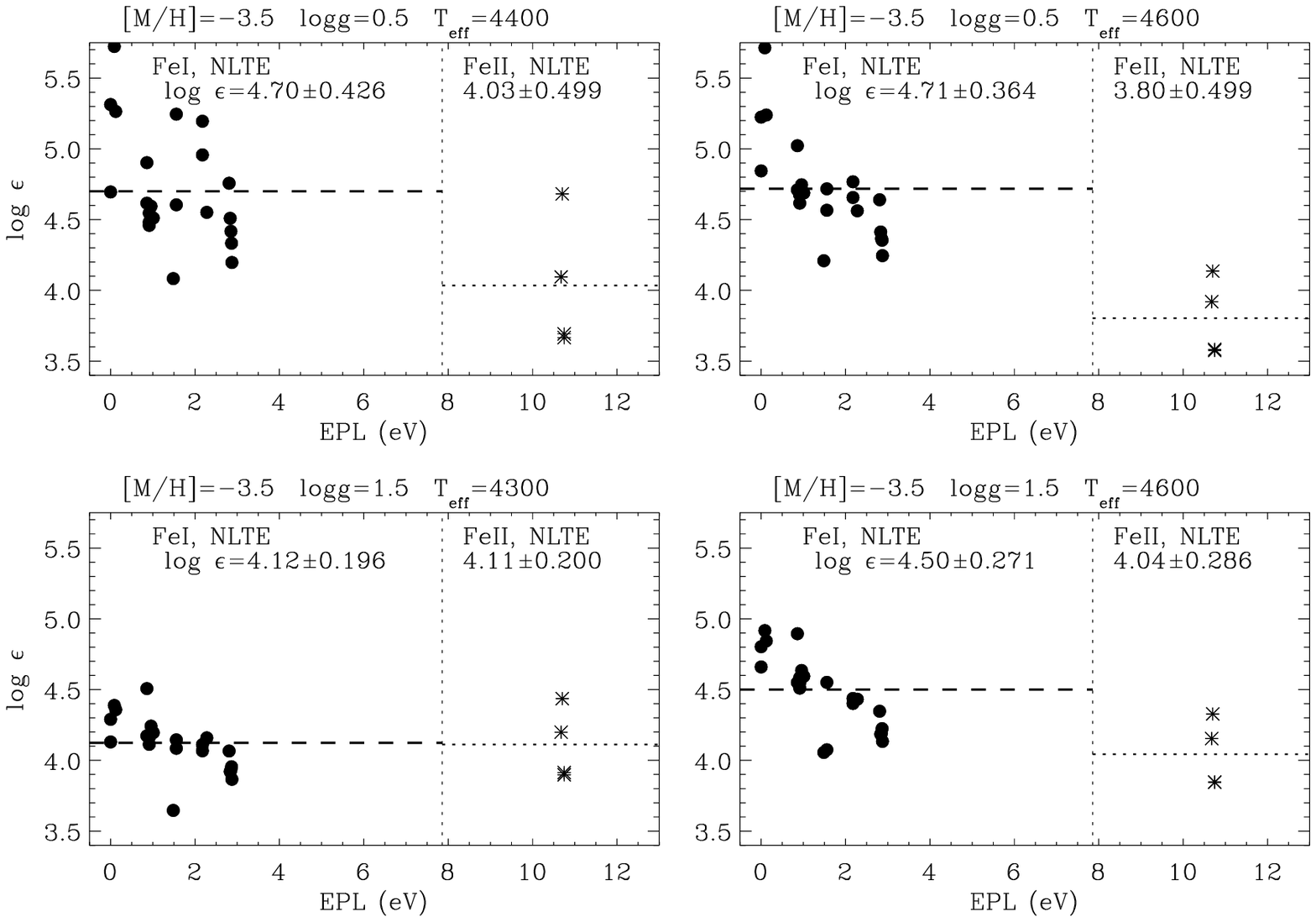,width=11cm,height=9.5cm,angle=360}
\caption[]{The same as in Fig. 4 but for a non-LTE Fe analysis.}
\end{figure*}

\begin{figure*}
\psfig{figure=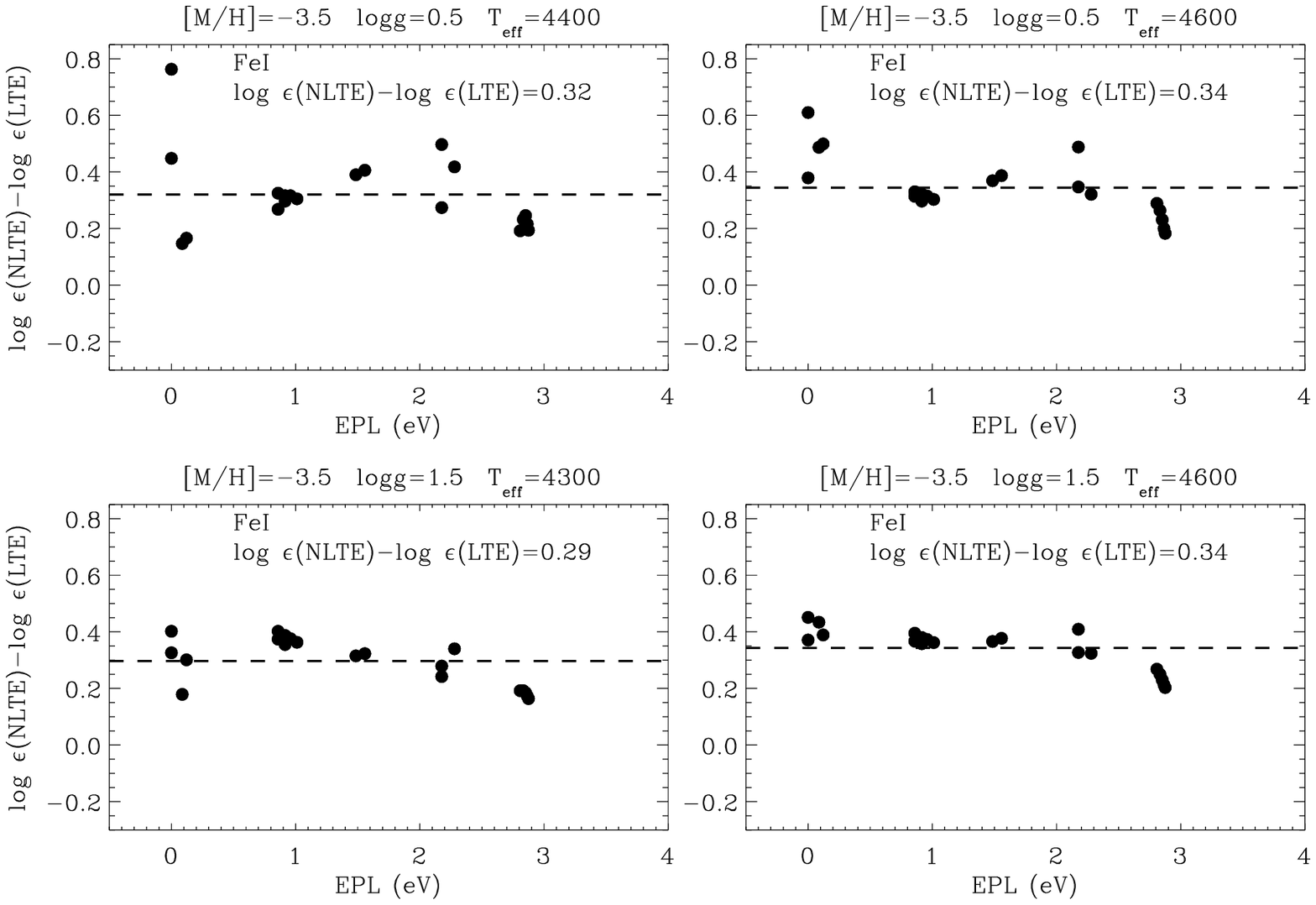,width=11.0cm,height=9.5cm,angle=360}
\caption[]{Difference between non-LTE and LTE abundances as a function of $\chi$ in CS\,29498--043.}
\end{figure*}

\def\baselinestretch{1}
\begin{table*}[t]
\caption[]{Equivalent widths of the oxygen and iron lines in CS\,22949--037. Data for the 
oxygen triplet comes from this work while the forbidden line 6300 \AA\ and the Fe lines 
are taken from Depagne et al. (2002). The wavelengths and the excitation energies of 
the Fe lines come from Moore (1959). The oscillator strengths of the \ion{Fe}{ii} and \ion{Fe}{i} 
were taken from Fuhr et al. (1988). The only exception are \ion{Fe}{i} 5049.82, 5232.946 (O'Brian et al.
1991) and 5324.185 \AA\ (Holweger et al. 1991).} 
\begin{scriptsize}
\begin{tabular}{lccccl}
\hline
\noalign{\smallskip}
Ion             & 
$\lambda$ (\AA) &
$\log gf$       &
$\chi$          &
EW             \\ 
\hline   
O\,{\sc i}   & 7771.960 & 0.324    & 9.11 & 41.0  \\
O\,{\sc i}   & 7774.180 & 0.174    & 9.11 & 30.0  \\
O\,{\sc i}   & 7775.400 & $-$0.046 & 9.11 & 19.0  \\
O\,{\sc i}   & 6300.230 & $-$9.759 & 0.00 &  5.0  \\
Fe\,{\sc i}  & 3899.709 & $-$1.531 & 0.09 & 102.2  \\
Fe\,{\sc i}  & 3920.260 & $-$1.746 & 0.12 &  95.5 \\
Fe\,{\sc i}  & 3922.914 & $-$1.651 & 0.05 &  102.5 \\
Fe\,{\sc i}  & 4005.246 & $-$0.610 & 1.55 &  61.5 \\
Fe\,{\sc i}  & 4045.815 & 0.280    & 1.48 &  99.6 \\
Fe\,{\sc i}  & 4063.597 & 0.070    & 1.55 &  89.8 \\
Fe\,{\sc i}  & 4071.740 & $-$0.022 & 1.60 &  84.1 \\
Fe\,{\sc i}  & 4076.636 & $-$0.360 & 3.20 &  6.3 \\
Fe\,{\sc i}  & 4132.060 & $-$0.648 & 1.60 &  58.2 \\
Fe\,{\sc i}  & 4143.871 & $-$0.450 & 1.55 &  66.2 \\
Fe\,{\sc i}  & 4147.673 & $-$2.104 & 1.48 &   7.7 \\
Fe\,{\sc i}  & 4181.758 & $-$0.180 & 2.82 &  10.5 \\
Fe\,{\sc i}  & 4187.044 & $-$0.548 & 2.44 &  18.6 \\
Fe\,{\sc i}  & 4199.098 & 0.250    & 3.03 &  23.9 \\
Fe\,{\sc i}  & 4202.031 & $-$0.708 & 1.48 &  60.3 \\
Fe\,{\sc i}  & 4222.219 & $-$0.967 & 2.44 &   8.3 \\
Fe\,{\sc i}  & 4227.434 & 0.272    & 3.32 &  15.1 \\
Fe\,{\sc i}  & 4250.125 & $-$0.405 & 2.46 &  22.6 \\
Fe\,{\sc i}  & 4260.479 & $-$0.020 & 2.39 &  44.2 \\
Fe\,{\sc i}  & 4271.159 & $-$0.349 & 2.44 &  32.7 \\
Fe\,{\sc i}  & 4282.406 & $-$0.810 & 2.17 &  16.0 \\
Fe\,{\sc i}  & 4325.765 & $-$0.010 & 1.60 &  91.2 \\
Fe\,{\sc i}  & 4404.752 & $-$0.142 & 1.55 &  85.1  \\
Fe\,{\sc i}  & 4415.125 & $-$0.615 & 1.60 &  61.8 \\
Fe\,{\sc i}  & 4447.722 & $-$1.342 & 2.21 &   6.0 \\
Fe\,{\sc i}  & 4461.654 & $-$3.210 & 0.09 &  33.0  \\
Fe\,{\sc i}  & 4528.619 & $-$0.822 & 2.17 &  21.3  \\
Fe\,{\sc i}  & 4871.323 & $-$0.410 & 2.85 &  10.7  \\
Fe\,{\sc i}  & 4872.144 & $-$0.600 & 2.87 &   7.8  \\
Fe\,{\sc i}  & 4891.496 & $-$0.140 & 2.84 &  19.8  \\
Fe\,{\sc i}  & 4920.509 & 0.060    & 2.82 &  23.5  \\
Fe\,{\sc i}  & 4994.133 & $-$3.080 & 0.91 &   7.5  \\
Fe\,{\sc i}  & 5001.871 & $-$0.010 & 3.87 &   3.1  \\
Fe\,{\sc i}  & 5049.825 & $-$1.355 & 2.27 &   5.5  \\
Fe\,{\sc i}  & 5051.636 & $-$2.795 & 0.91 &  11.2  \\
Fe\,{\sc i}  & 5068.774 & $-$1.230 & 2.93 &   3.9  \\
Fe\,{\sc i}  & 5110.414 & $-$3.760 & 0.00 &  15.9  \\
Fe\,{\sc i}  & 5123.723 & $-$3.068 & 1.01 &   4.8  \\
Fe\,{\sc i}  & 5127.363 & $-$3.307 & 0.91 &   3.0  \\
Fe\,{\sc i}  & 5166.286 & $-$4.195 & 0.00 &   7.8  \\
Fe\,{\sc i}  & 5171.599 & $-$1.793 & 1.48 &  19.5  \\
Fe\,{\sc i}  & 5194.943 & $-$2.090 & 1.55 &   6.9  \\
Fe\,{\sc i}  & 5232.946 & $-$0.057 & 2.93 &  15.6  \\
Fe\,{\sc i}  & 5266.562 & $-$0.490 & 2.99 &   6.2  \\
Fe\,{\sc i}  & 5324.185 & $-$0.100 & 3.20 &   7.2  \\
Fe\,{\sc i}  & 5339.935 & $-$0.680 & 3.25 &   3.2  \\
Fe\,{\sc i}  & 5371.493 & $-$1.645 & 0.95 &  62.6  \\
Fe\,{\sc i}  & 5383.374 & 0.500    & 4.29 &   2.3  \\
Fe\,{\sc i}  & 5397.131 & $-$1.993 & 0.91 &  46.8  \\
Fe\,{\sc i}  & 5405.778 & $-$1.844 & 0.99 &  44.8  \\
Fe\,{\sc i}  & 5429.699 & $-$1.879 & 0.95 &  48.5  \\
Fe\,{\sc i}  & 5434.527 & $-$2.122 & 1.01 &  31.5  \\
Fe\,{\sc i}  & 5446.920 & $-$1.930 & 0.99 &  43.8  \\
Fe\,{\sc i}  & 5506.782 & $-$2.797 & 0.99 &   9.4  \\
Fe\,{\sc ii} & 4178.85 & $-$2.480 & 2.57 & 5.0  \\
Fe\,{\sc ii} & 4233.16 & $-$2.000 & 2.57 & 20.2  \\
Fe\,{\sc ii} & 4416.81 & $-$2.600 & 2.77 & 4.9  \\
Fe\,{\sc ii} & 4515.33 & $-$2.480 & 2.83 & 3.4  \\
Fe\,{\sc ii} & 4520.22 & $-$2.600 & 2.79 & 2.8  \\
Fe\,{\sc ii} & 4555.89 & $-$2.290 & 2.82 & 7.3  \\
\noalign{\smallskip}
\hline
\end{tabular}
\end{scriptsize}
\label{tab2}
\end{table*}
\def\baselinestretch{2}

\section{Stellar parameters from the non-LTE computations of iron}

It is well known that a strong over-ionization of neutral iron in metal-poor stars 
leads to the systematic difference in abundances determined from the Fe\,{\sc i} and 
Fe\,{\sc ii} lines. This difference increases with decreasing metallicity and may reach
0.4 dex in very metal-poor dwarfs (Th\'evenin \& Idiart 1999). In addition, 
the non-LTE modeling predicts a dependence of the non-LTE abundance corrections  
of Fe\,{\sc i} lines on the lower excitation potential ($\chi$). The corrections are
particularly large for the low-excitation Fe\,{\sc i} lines, while the non-LTE effects 
are not important for the Fe\,{\sc ii} lines. The non-LTE abundance 
corrections in the Sun are in the range 0.02--0.1~dex (Shchukina \& Trujillo Bueno 2001).
In general, non-LTE effects play a significant role in  metal-poor stars because of 
decrease in electron density when collisions with free electrons no longer dominate the
kinetic equilibrium. Another consequence of the metal deficiency is an appreciable 
weakening in UV blanketing. Non-LTE over-ionization of Fe\,{\sc i} substantially reduces the
UV line opacity and allows more flux to escape. These effects lead us to suspect that 
the gravities of metal-poor giants derived from the LTE Fe analysis are underestimated 
because of the neglect of non-LTE effects. This effect has been studied in metal-poor 
dwarfs (Th\'evenin \& Idiart 1999) and in the subgiant BD +23\,1330 
(Israelian et al.\,2001). 
In fact, Th\'evenin \& Idiart (1999) derived gravity 
corrections of up to 0.5 dex with respect to LTE values, for the case of stars with 
[Fe/H] $\sim -3.0$. They have shown that non-LTE effects are important
in determining stellar parameters from the iron ionization balance.

The stellar parameters of our targets were obtained using the ionization equilibrium of Fe. 
The Fe model atom used in our study provides very consistent results for a 3D atmospheric 
model of the Sun (Shchukina \& Trujillo Bueno 2001). The microturbulent velocity was fixed
at 2 \hbox{km$\;$s$^{-1}$}.
Non-LTE analysis of Fe based on plane--parallel atmosphere models of Kurucz (1992)  was 
carried out with the code NATAJA  (Shchukina \& Trujillo Bueno 2001), and the atmospheric 
parameters were derived using the same method as in Israelian et al.\,(2001). The equivalent 
widths of 26 (CS\,29498--043) and 60 (CS\,22949--037) Fe lines listed in Tables 1 and 2 were
taken from the articles of Aoki et al.\,(2002) and Depagne et al.\,(2002), respectively. 
In the present analysis we used solar abundances from Grevesse and Sauval (1998). The only
exception was oxygen, for which we used the solar abundance $\log \epsilon(O)$=8.74 from 
Nissen et al. (2002).

Figures 1 to 6 show our computations for a grid of atmospheric models for CS\,22949-037 and
CS\,29498-043. The non-LTE corrections to [Fe/H] are around 0.3--0.4 dex (Figs 3 and 6) and
abundance scatter obtained for the best parameter sets are less than 0.2 dex  (Figs 2 and 5).  
After many iterations with different input parameters, our final results were 
$T_{\rm eff}$ = 4900$\pm$125 K, $\log g$ = 2.5$\pm$0.3 and [Fe/H ] = $-$3.5$\pm$0.2 
for CS\,22949--037 and $T_{\rm eff}$ = 4300$\pm$160 K, $\log g$ = 1.5$\pm$0.35 and 
[Fe/H] = $-3.5\pm0.24$ for CS\,29498--043. The errors were computed following McWilliam et al. (1995)
and assuming those authors 1-$\sigma$ uncertainties of $\sim$0.03 and $\sim$0.07 dex for 
the oscillator strengths of the \ion{Fe}{i} and \ion{Fe}{ii} lines, respectively. We also note that the
oscillator strengths of the \ion{Fe}{ii} lines used in our calculations (Fuhr et al. 1988) are very 
similar to those compiled by McWilliam et al. (1995). The gravities that we obtained are about 1\,dex 
larger compared with those reported by Aoki et al.\ (2002) and Depagne et al.\,(2002).

\def\baselinestretch{1}
\begin{table}[t]
\caption[]{Equivalent widths of the oxygen and iron lines in CS\,29498--043. 
Oxygen lines are measured by us while the source of Fe the lines is the paper 
by Aoki et al. (2002). The wavelengths and the excitation energies of 
the Fe lines come from Moore (1959). The oscillator strengths of the \ion{Fe}{ii} and \ion{Fe}{i} 
were taken from Fuhr et al. (1988) except \ion{Fe}{i} 5049.825 (O'Brian et al. 1991) and
4957.603 \AA\ (Gigas 1988).}
\begin{scriptsize}
\begin{tabular}{lccccl}
\hline
\noalign{\smallskip}
Ion             & 
$\lambda$ (\AA) &
$\log gf$       &
$\chi$          &
EW             \\
\hline
O\,{\sc i}   & 7771.960 & 0.324    & 9.11 & 18.0  \\
O\,{\sc i}   & 7774.180 & 0.174    & 9.11 & 15.0 \\
O\,{\sc i}   & 7775.400 & $-$0.046 & 9.11 & 10.0 \\
O\,{\sc i}   & 6300.230 & $-$9.759  & 0.00 &  60. \\
Fe\,{\sc i}  & 4005.246 & $-$0.610 & 1.55 & 87.7 \\
Fe\,{\sc i}  & 4459.121 & $-$1.279 & 2.17 & 33.1  \\
Fe\,{\sc i}  & 4461.654 & $-$3.210 & 0.09 & 83.2  \\
Fe\,{\sc i}  & 4489.741 & $-$3.966 & 0.12 & 43.3  \\
Fe\,{\sc i}  & 4528.619 & $-$0.822 & 2.17 & 52.1  \\
Fe\,{\sc i}  & 4890.762 & $-$0.430 & 2.86 & 27.6  \\
Fe\,{\sc i}  & 4891.496 & $-$0.140 & 2.84 & 46.0  \\
Fe\,{\sc i}  & 4918.999 & $-$0.370 & 2.85 & 35.4  \\
Fe\,{\sc i}  & 4920.509 & 0.060    & 2.82 & 56.7  \\
Fe\,{\sc i}  & 4939.690 & $-$3.340 & 0.86 & 36.0  \\
Fe\,{\sc i}  & 4957.603 & 0.043    & 2.80 & 67.4  \\
Fe\,{\sc i}  & 4994.133 & $-$3.080 & 0.92 & 27.0  \\
Fe\,{\sc i}  & 5012.071 & $-$2.642 & 0.86 & 55.8  \\
Fe\,{\sc i}  & 5049.825 & $-$1.355 & 2.28 & 27.5  \\
Fe\,{\sc i}  & 5051.636 & $-$2.795 & 0.91 & 43.6  \\
Fe\,{\sc i}  & 5083.342 & $-$2.958 & 0.96 & 36.8  \\
Fe\,{\sc i}  & 5110.414 & $-$3.760 & 0.00 & 65.9  \\
Fe\,{\sc i}  & 5123.723 & $-$3.068 & 1.01 & 26.6  \\
Fe\,{\sc i}  & 5127.363 & $-$3.307 & 0.91 & 21.3  \\
Fe\,{\sc i}  & 5166.286 & $-$4.195 & 0.00 & 36.5  \\
Fe\,{\sc i}  & 5171.599 & $-$1.793 & 1.48 & 34.4  \\
Fe\,{\sc i}  & 5194.943 & $-$2.090 & 1.55 & 39.9  \\
Fe\,{\sc ii} & 4522.634 & $-$2.030 & 2.83 & 28.0  \\
Fe\,{\sc ii} & 4583.829 & $-$2.020 & 2.79 & 25.9  \\
Fe\,{\sc ii} & 4923.921 & $-$1.320 & 2.88 & 50.1  \\
Fe\,{\sc ii} & 5018.434 & $-$1.220 & 2.88 & 60.3  \\
\noalign{\smallskip}
\hline
\end{tabular}
\end{scriptsize}
\label{tab2}
\end{table}
\def\baselinestretch{2}

\begin{figure*}
\psfig{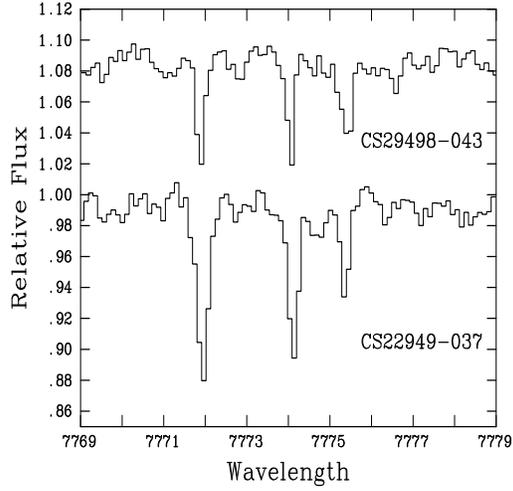}
\caption[]{The oxygen triplet observed in CS\,29498--043 (shifted upwards by 0.08) 
and CS\,22949--037.}
\end{figure*}

\begin{figure*}
\psfig{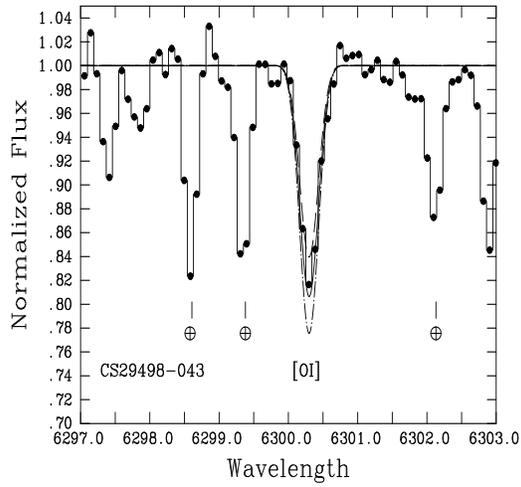}
\caption[]{The oxygen line at 6300 \AA\ observed in CS\,29498--043. Spectrum synthesis was
carried out for the values [O/H] = $-$1.1 (dashed), $-$1.0 (solid), $-$0.9 (dashed-dotted) and assuming
$v\sin i$=8\,km\,s$^{-1}$, $V_{\rm macro}$=2 km\,s$^{-1}$ and 0.6 for the 
limb-darkening coefficient.}
\end{figure*}

\section{Oxygen}

We have detected all three lines of the oxygen triplet in CS\,22949--037 and CS\,29498--043 (Fig. 7)
and the [O\,{\sc i}] 6300 \AA\ line in CS\,29498-043 (Fig. 8). 
The equivalent width of the forbidden line in CS\,29498-043 EW$\sim$35 m\AA\ measured by 
Aoki et al.\,(2003) using SUBARU/HDS is smaller than the value obtained from our 
TNG/SARG. Our TNG/SARG measurement
agrees within 3-$\sigma$. This modest agreement is not surprising since the line detected by
Aoki et al.\,(2003) is strongly blended with a telluric feature and our line was affected by
sky emission. The relatively large difference may then be related to uncertainties in the
corrections for telluric absorption and/or sky emission.  
The non-LTE computations of the oxygen atom were carried out using the atomic model with 23 levels
of O\,{\sc i} and one level of O\,{\sc ii} described by Carlsson \& Judge (1993). While only
31 bound--bound and 23 bound--free radiative transitions were considered in our computations,
we note that the consideration of additional levels and transitions does not affect our results 
(Shchukina 1987; Takeda 2003). Our computations for the Sun and metal-poor dwarfs
predict non-LTE corrections that are very close to those reported recently by Takeda (2003) 
and Nissen et al. (2002) for 1D models. 

It is well known that inelastic collisions with hydrogen atoms tend to cancel out the non-LTE
effects. The forbidden line is not affected by these collisions since it is formed under LTE. 
The effect on the triplet may reach 0.1 dex in hot and metal-poor subdwarfs such as LP815-43 
(e.g. Nissen et al. 2002). However, it is often stated that Drawin's formalism (Drawin 1968) gives 
very uncertain results for hydrogen collision rates (e.g. Belyaev et al. 1999). Obviously, 
we are not concerned with these problems in our targets since the density of H atoms (and therefore
the collision rates) in the atmosphere of K giants is about two orders of magnitude lower than in 
dwarfs. Moreover, inelastic collisions with hydrogen atoms bring the oxygen abundance derived from
the 7771-5 \AA\ triplet close to its LTE value. This will make the ``oxygen conflict'' even more 
severe as the discrepancy between the triplet and the forbidden line will be greater. Thus, collisions 
with H atoms were not taken into account in our computations.
We also note that the effects of triplet--quintet system coupling, CO-binding and $L_{\beta}$
pumping are negligible in the atmospheres of our targets. 

Our non-LTE oxygen abundances from the near-IR triplet yield [O/Fe] = 3.13$\pm$0.21 and 
[O/Fe] = 3.02$\pm$0.27 in CS\,22949--037 and CS\,29498--043, respectively. The non-LTE 
corrections ($\Delta \epsilon=\epsilon({\rm non-LTE})-\epsilon({\rm LTE})$) on the triplet 
and 6300 \AA\ are listed in Table 5. Assuming our stellar parameters for CS\,22949--037  
and the equivalent width of the forbidden line from Depagne et al.\,(2002) we find
[O/Fe] = 1.95. The large difference between the abundances derived from the triplet and the 
forbidden line cannot be explained by the non-LTE abundance corrections for the triplet (Table 4)
and/or by a noise/telluric correction for the forbidden line. In fact, one needs [O\,{\sc i}] 
6300 \AA\ line with an EW $>$ 50 m\AA\ to provide [O/Fe] $>$ 3. This is clearly ruled out by 
observations of Depagne et al. (2002). Even if we assume for CS\,22949--037 the stellar parameters 
from Depagne et al. (2002), the difference between 7771--5 and 6300 \AA\ is as high as 1.55 dex.
The forbidden line  measured in the SARG spectra of CS\,29498--043 has EW = 60$\pm$10m\AA, 
providing [O/Fe] = 2.49$\pm$0.13 (Fig. 8).

It is hard to resolve the conflict between the triplet and the forbidden line by 
playing with the stellar parameters since the errors in $T_{\rm eff}$ and $\log g$
are of the order of 150 K and 0.3 dex, respectively. We have to understand
the formation mechanisms of these lines in the atmospheres of our targets.  
The intensity contribution functions (CFI) (see Gray 1976) of the triplet and the forbidden
lines decrease rapidly with atmospheric height (Fig. 9), and one can assume, as a working hypothesis,
that these lines are optically thin. In this case
their equivalent widths will be proportional to the ratio $K^{\rm line}$ / $K^{\rm cont}$, where 
$K^{\rm line}$ and $K^{\rm cont}$ are the absorption coefficients in the line and in the continuum,
respectively. Many authors state 
that the strength of the forbidden line is sensitive to the structure of the upper atmosphere. 
Our analysis (Fig. 9) shows that this is not the case since the forbidden line has a very small 
oscillator strength and is therefore formed deep in the atmosphere. The triplet lines are 
formed at this depth because they are excited from high excitation levels which are populated
only in the hot layers of the atmosphere. The absorption coefficient in the line will be very
sensitive to $T_{e}$ since the population of the lower level is proportional to 
exp($- \chi_{l}$\,/\,$kT_{e}$) and therefore EW$_{7771-5}$ $\sim$ $T{_e}$. The strength of the 
forbidden line will be controlled by $K^{\rm cont}$ because $K^{line} \ll K^{\rm cont}$. 
Since $K^{\rm cont} \sim N{_e}$ (where $N{_e}$ is 
the electron density), EW$_{6300}$ $\sim$ 1\,/\,$N{_e}$. Thus, the EW$_{6300}$ will decrease
(while EW$_{7771-5}$ will increase) with an increasing $T{_e}$ since $N{_e}$ $\sim$ $T{_e}$ in 
stellar atmospheres. Given the EWs of the triplet and the forbidden line, one may find  the 
effective temperature of the model where these two indicators provide the same abundance. This test has 
been carried out for CS\,22949--037 and the results are shown in Fig.\,10. While the $T_{\rm eff}$ 
of the star derived from the Fe lines is 4900 K, consistent oxygen abundance can only be obtained 
when $T_{\rm eff}$ = 5600 K.

The temperature distributions in two stars with $T_{\rm eff}$ = 4900 and 5600 K and the location 
of the maximum CFI in the line centre for each model are shown in the Figs.\,11 and 12. These figures 
clearly demonstrate that the formation depth of the forbidden line moves toward the surface and 
to much higher temperatures in the model with a higher $T_{\rm eff}$. In fact, the maximum contribution 
to the line strength comes from the hot layers (Fig.\,11). The same plot for the near-IR triplet shows 
(Fig.\,12) that the maximum contribution to the equivalent width comes from the layers that are closer 
to the surface compared with the forbidden line. In fact, our analysis demonstrates that the formation 
layer of the forbidden line is optically thinner and therefore less sensitive to the temperature 
gradient (i.e. $T_{\rm eff}$). Our analysis shows that the failure of the Kurucz (1992) models to provide
consistent abundances from the near-IR triplet and the forbidden line, comes from the temperature and 
density gradients near the continuum-forming region. This problem has the same roots as that
related to the discrepancy between the temperatures obtained from colours (continuum) and the
Fe ionization balance. It is clear that we cannot resolve the conflict between the triplet and 
the forbidden line by modifying the temperature distribution in the upper atmosphere. 
The key to this problem is hidden deep in the atmosphere. There must be a certain
combination of the electron density and the temperature gradients close to the continuum 
formation region to provide consistent abundances from the triplet and the forbidden line.
 It is easy to show that the forbidden line is formed deep in the atmosphere of our targets.
We have repeated calculations of this line for CS\,22949-037 by gradually removing the upper 
parts of the atmosphere until we reached those layers where the 6300 \AA\ line is actually ``formed''. 
These simple tests have demonstrated that more than 80\,\% of the EW of the forbidden line is formed at 
$\log {\it mass} >$ 1.4 (e.g. Fig.\,9).
The discrepancy between the near-IR triplet and the 6300 \AA\ line also exists in moderately
metal-poor giants. These problems have been addressed in detail by Takeda (2003) and
Fulbright \& Johnson (2003). Our targets provide the most extreme cases of the ``oxygen conflict'' 
at very low metallicities. 

We have found that the discrepancy between the near-IR triplet and the forbidden line cannot be
explained by non-LTE effects, uncertainties in the stellar parameters or the quality
of observations. The oxygen puzzle arises from the failure of  standard plane-parallel atmosphere 
 to describe physical conditions in the atmospheres of very metal-poor giants.
Spherical effects cannot be the cause of this discrepancy since these lines are formed in
very deep and narrow layers where the plane--parallel approximation is certainly applicable. 
The oxygen puzzle is not unique to CS\,22949--037 and CS\,29498-043. On a smaller scale it exists 
in other halo giants (Takeda 2003; Fulbright \& Johnson 2003). We suspect that a similar discrepancy 
exists for the most metal-poor star in our Galaxy, HE0107-5240.

\def\baselinestretch{1}
\begin{table}[t]
\caption[]{Non-LTE abundance of oxygen and magnesium lines in CS\,29498--043 derived for three sets of the
atmospheric parameters. Each set is the best model obtained for a given [Mg/Fe] ratio provided on the second
line. The second column gives the EWs of the Mg and O lines from Aoki et al. (2002) and this work, respectively.
}
\begin{scriptsize}
\begin{tabular}{lcccc}
\hline
\noalign{\smallskip}
Line               & 
EW                 &
(4400/0.5/$-$3.5)  &
(4400/1.5/$-$3.5)  &
(4300/1.5/$-$3.5)  \\
 & m\AA\ & [Mg/Fe] = 1.8 &  [Mg/Fe] = 1.0 & [Mg/Fe] =  0 \\
\hline
Mg\,{\sc i} 4571  &  104     &  	7.286	&	 6.233  &	  5.626 \\
Mg\,{\sc i} 5172  &  218     &  	5.031	&	 4.905  &	  5.147 \\
Mg\,{\sc i} 5183  &  228     &  	4.937	&	 4.785  &	  5.012 \\
\hline
O\,{\sc i}  6300  &  60     &		 7.626  &	 7.732  &	 7.694  \\
O\,{\sc i}  7772  &  18     &		 7.984  &	 8.232  &	 8.296	\\
O\,{\sc i}  7774  &  15     &		 8.019  &	 8.274  &	 8.343	\\
O\,{\sc i}  7775  &  10     &		 8.004  &	 8.272  &	 8.345  \\
\noalign{\smallskip}
\hline
\end{tabular}
\end{scriptsize}
\end{table}

\begin{table}[t]
\caption{Non-LTE abundance of oxygen and magnesium in CS\,22949--037 derived from the model
with $T_{\rm eff}$ = 4900 K, $\log g$ = 2.5 and [Fe/H] = $-$3.5. The last column gives the 
abundance correction due to non-LTE effects. The observed EWs of Mg and [O\,{\sc i}] 6300 \AA\ were 
taken from Depagne et al. (2002).}
\begin{scriptsize}
\begin{tabular}{lccc}
\hline
\noalign{\smallskip}
Line                        & 
EW (m\AA)                   &
log $\epsilon$             &
$\Delta \epsilon({\rm NLTE-LTE})$               \\
\hline
Mg\,{\sc i} 3829  &  156.1  &      5.224	&    0.187 \\
Mg\,{\sc i} 3832  &  185.2  &      4.986	&    0.158 \\
Mg\,{\sc i} 3838  &  202.7  &      4.874	&    0.160 \\
Mg\,{\sc i} 4571  &   52.9  &      5.810	&    0.670 \\
Mg\,{\sc i} 5172  &  176.1  &      5.347	&    0.141 \\
Mg\,{\sc i} 5183  &  199.4  &      5.353	&    0.124 \\
\hline
O\,{\sc i}  6300  & 5.   &  7.184	&      $-$0.013  \\
O\,{\sc i}  7772  & 41.  &  8.430	&      $-$0.238  \\
O\,{\sc i}  7774  & 30.  &  8.362	&      $-$0.222  \\
O\,{\sc i}  7775  & 19.  &  8.303	&      $-$0.200  \\
\noalign{\smallskip}
\hline
\end{tabular}
\end{scriptsize}
\end{table}

\begin{table*}[t]
\caption{Non-LTE abundance corrections to the oxygen and magnesium lines.
$\Delta_{\rm Mg(Sum)}$ 
for all Mg lines except 4571 \AA\ listed in Tables 3 and 4. The last two columns provide the abundance differences
derived from the triplet and the forbidden line ($\Delta A_{O}$) and from the 4571 \AA\ and the 
remaining Mg\,{\sc i} lines from Tables 3 and 4.} 
\begin{scriptsize}
\begin{tabular}{lcccccc}
\hline
\noalign{\smallskip}
Target                         & 
$\Delta_{\rm [OI]}$            &
$\Delta_{\rm Triplet}$         &
$\Delta_{\rm Mg 4571}$  &
$\Delta_{\rm Mg(Sum)}$   &
$\Delta A_{\rm O}$              &
$\Delta A_{\rm Mg}$              \\  
\hline
CS\,29498--043 & $-$0.012 & $-$0.188 & 0.179 & $-$0.071 & 0.53    & 0.546 \\
CS\,22949--037 & $-$0.013 & $-$0.22  & 0.670 &    0.154 & 1.18    & 0.653 \\
\noalign{\smallskip}
\hline
\end{tabular}
\end{scriptsize}
\end{table*}


\begin{figure*}
\psfig{figure=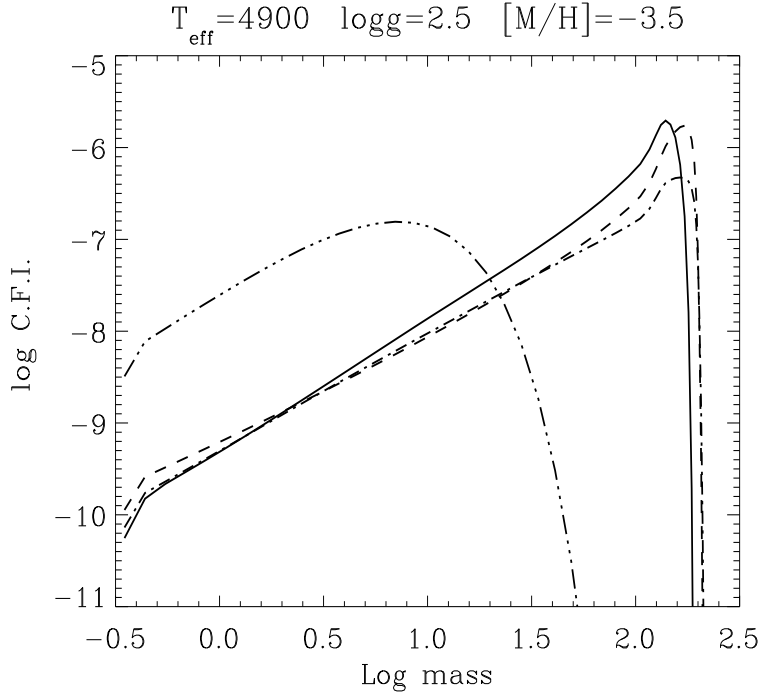,width=11cm,height=9.5cm,angle=360}
\caption[]{The intensity contribution functions (CFI) for the line centre in the 
atmosphere of CS\,22949--037 computed for the forbidden line [O\,{\sc i}] 6300 \AA\ 
(dashed line), the near-IR triplet (solid line), the Mg\,{\sc i} lines 4571 \AA\ 
(dashed-dotted) and 5183 \AA\ (dashed-three dotted). The log mass is above 1 cm$^2$.}
\end{figure*}

\begin{figure*}
\psfig{figure=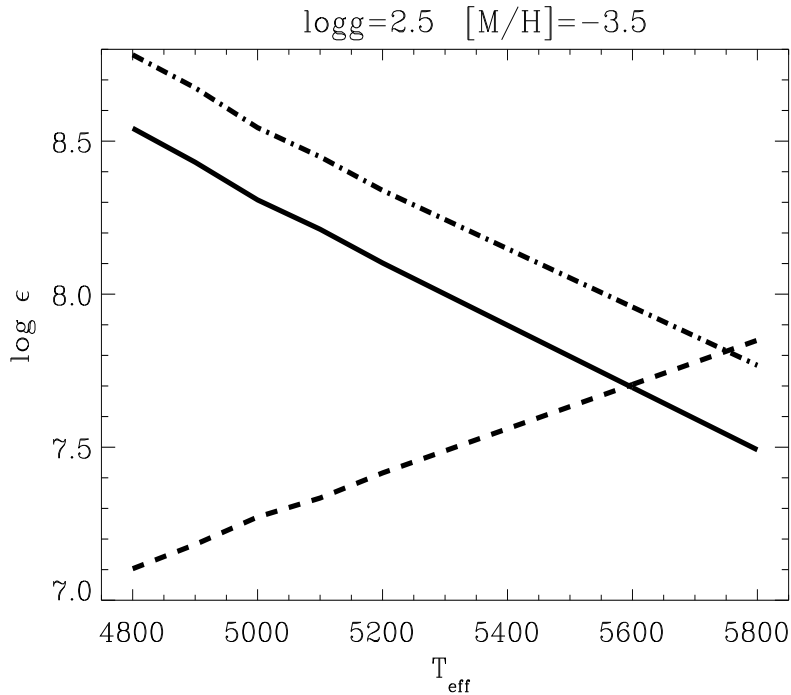,width=11cm,height=9.5cm,angle=360}
\caption[]{The oxygen abundance derived from the triplet (non-LTE is a solid line and LTE is dashed-dotted) and
the forbidden line (dashed) in CS\,22949--037 as function of $T_{\rm eff}$. The EWs were
taken from Table 1.}
\end{figure*}

\begin{figure*}
\psfig{figure=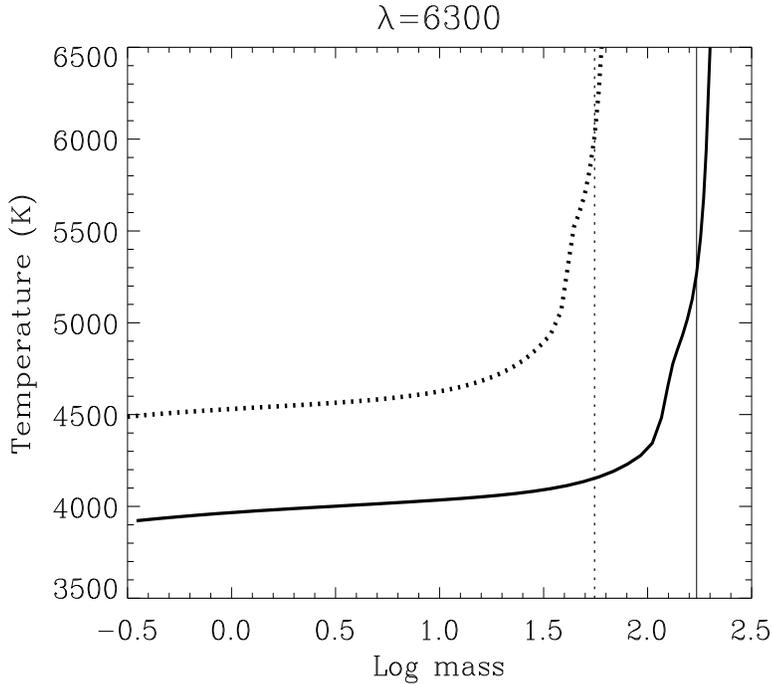,width=11cm,height=9.5cm,angle=360}
\caption[]{The temperature distribution for two models with $T_{\rm eff}$ = 5600 K (dotted line) and
$T_{\rm eff}$ = 4900 K (solid line). The dotted and solid vertical lines show the location of
the maximum CFI for the forbidden line [O\,{\sc i}] 6300 \AA. The gravity and the metallicity are
set at $\log g$ = 2.5 and [Fe/H] = $-$3.5, respectively. See 
Section 4 for details. The log mass is above 1 cm$^2$.}
\end{figure*}

\begin{figure*}
\psfig{figure=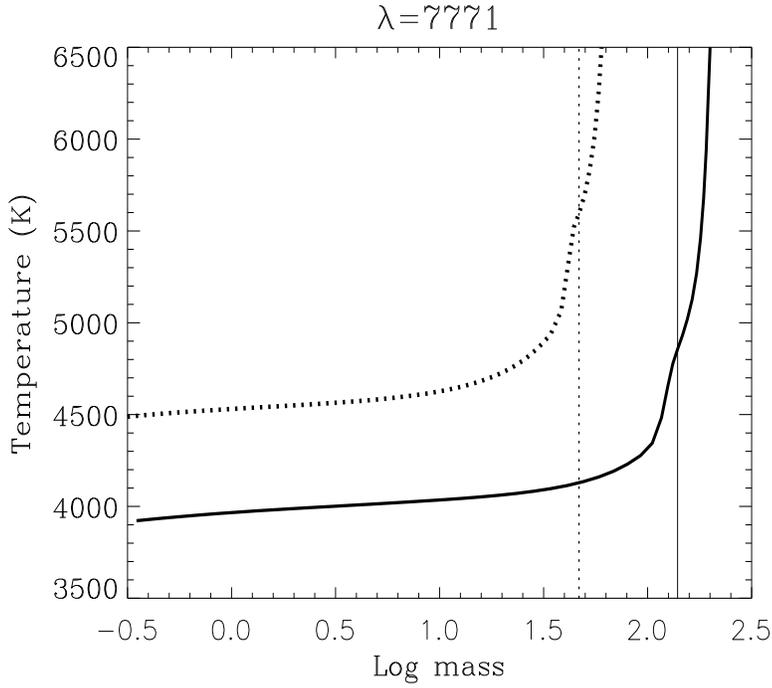,width=11cm,height=9.5cm,angle=360}
\caption[]{The same as in Fig. 11 but for the near-IR triplet. The log mass is above 1 cm$^2$.}
\end{figure*}

\begin{figure*}
\psfig{figure=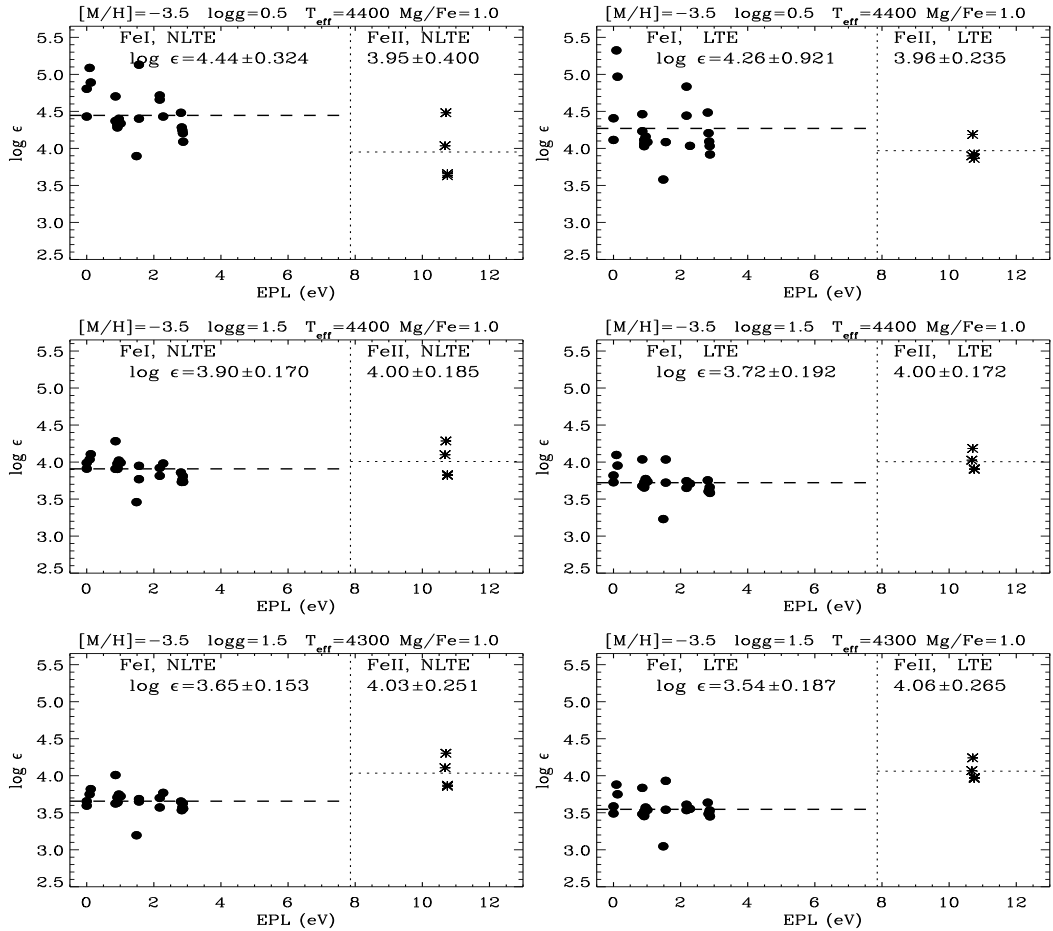,width=14cm,height=12.5cm,angle=360}
\caption[]{Results of LTE and non-LTE Fe abundance determination for the Fe\,{\sc i} and Fe\,{\sc ii} lines 
in CS\,29498--043 for a small grid of model atmospheres and for [Mg/Fe] = 1.}
\end{figure*}

\begin{figure*}
\psfig{figure=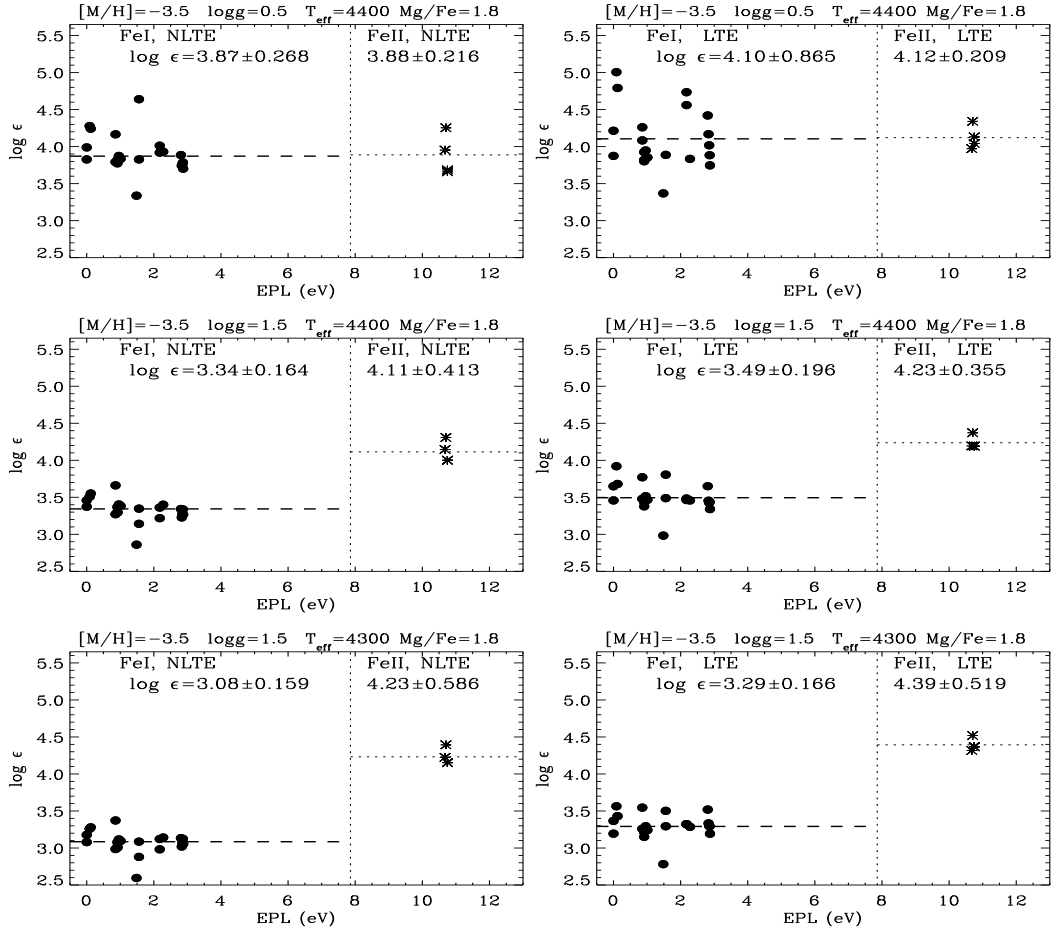,width=14cm,height=12.5cm,angle=360}
\caption[]{The same as in Fig. 13 but for [Mg/Fe] = 1.8.}
\end{figure*}

\begin{figure*}
\psfig{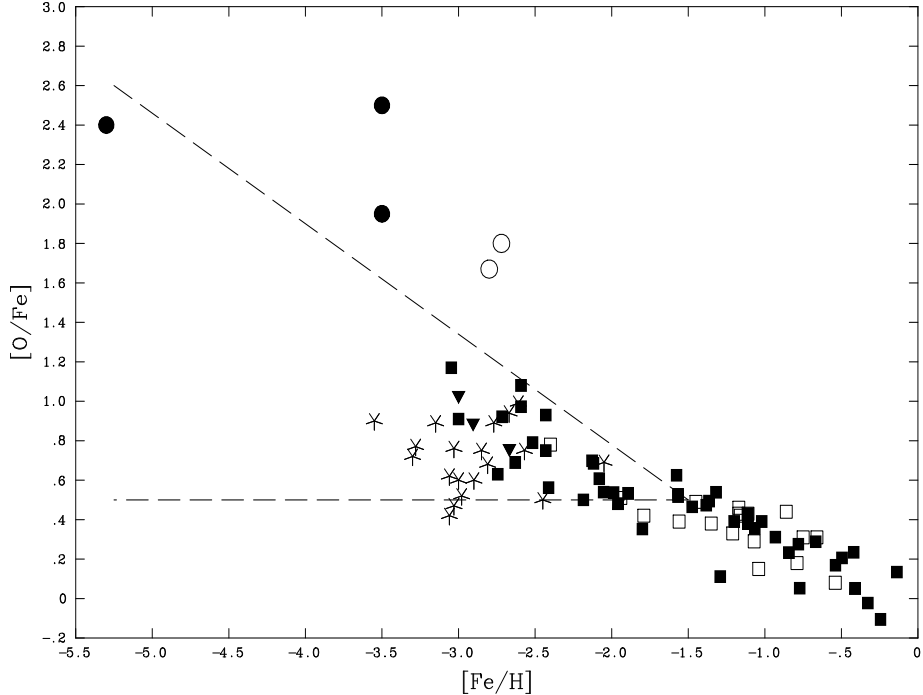}
\caption[]{Oxygen overabundance versus Fe. Data from Israelian et al.\,(2001), 
Cayrel et al.\,(2003), and Nissen et al.\,(2002) are marked by filled squares, 
stars, and open squares, respectively. Filled triangles represent the upper 
limits from Israelian et al.\,(2001). The values were not corrected for 3D effects 
because the latter do not predict consistent oxygen abundance from different 
indicators. We cannot incorporate these corrections in the
data of Cayrel et al.\,(2003) since there are no 3D models available for giants. 
Three filled circles indicate the oxygen abundance derived from the
forbidden line for the two giants discussed in this article and for
HE 0107-5240 derived from UV OH lines by Bessell et al. (2004).
The spectroscopic binaries CS\,29497-030 ([Fe/H]\,=\,$-$2.8) and LP625-44 ([Fe/H]\,=\,$-$2.72) 
with a large enhancement in s-process elements are indicated by empty circles and 
have [O/Fe]\,=\,1.67 (Sivarani et al. 2003) and [O/Fe]\,=\,1.8 (Aoki et al. 2002b), respectively.}
\end{figure*}

\section{Magnesium}

The non-LTE computations of the magnesium atom were carried out using the 24 levels 
simplified version of the model atom
described by Carlsson, Rutten \& Shchukina (1992). Addition of new levels and transitions 
does not influence on the solution for the lines considered in this paper.
The abundance of Mg in CS\,29498--043 was derived 
from the spectral lines at 4571, 5172 and 5183 \AA\  using the equivalent width measurements of
Aoki et al. (2002). These lines provide very different abundances just as in the case of 
oxygen (Table 3). The forbidden resonance line Mg\,{\sc i} 4571 \AA\ in CS\,29498--043 provides 
[Mg/Fe] = 1.626 with a non-LTE correction of $\sim$0.18 dex while Mg\,{\sc i} lines at 5172 and 
5183 \AA\ provided consistent abundance with a mean [Mg/Fe] = 1.08. The non-LTE corrections
in 4571 \AA\ are 0.18 dex while in two other lines they even do not reach $-$0.1 dex (Table 5).
The difference between the 4571 and 5172 + 5183 \AA\ is larger in the non-LTE than in the LTE case. 
The mean abundance from the three lines is [Mg/Fe] = 1.26 which, of course, does not makes any sense 
given the huge discrepancy between the 4571 and 5172 + 5183 \AA\ lines. As for the CS\,22949--037, 
there are six Mg\,{\sc i} lines available (Table 4) in our model atom from the article of Depagne et al.\,(2002). 
From the five lines listed in Depagne et al.\,(2002) we obtained a mean [Mg/Fe] = 1.156 while 
the 4571 \AA\ line again yields a much larger abundance [Mg/Fe] = 1.81. 

The CFIs of the 4571 and 5183 \AA\ lines in CS\,22949--037 have very different shapes (Fig. 9). The 
5183 \AA\ line is formed in an extended region of the upper atmosphere while the forbidden resonance line is 
produced by the same deep layers where the 6300 \AA\ forbidden line of the neutral oxygen is formed. The 
5183 \AA\ line is very strong and less sensitive to the effective temperature compared with the 
4571 \AA\ line. The agreement between abundances provided by 4571 \AA\ and 5183 \AA\ lines can 
be achieved if we increase the temperature in the inner
layers of the atmosphere. This will increase the strength of the 4571 \AA\ line (and the abundance obtained
from this line will decrease) while the 5183 \AA\ will remain almost unchanged. Given the negligible
sensitivity of the 5183 \AA\ line to the effective temperature one may suggest that this line provides a
more reliable abundance. This is similar to the oxygen atom, where the forbidden line is believed to 
be the best abundance indicator since it is less sensitive to $T_{\rm eff}$.

Mg is an important source of free electrons and its overabundance may have a non-negligible effect on 
the stellar surface gravity derived from the ionization balance of Fe. According Aoki et al. (2002),
the effect of Mg overabundance on the gravity of CS\,29498--043  is 0.4 dex. We have repeated
the non-LTE Fe and Mg analysis of CS\,29498--043 for the cases when [Mg/Fe] = 1.0 and [Mg/Fe] = 1.8 
(see Figs 13 and 14).
It appears that the gravity of the star is not changed when the Mg abundance is increased by a factor of 10  
(i.e. from [Mg/Fe] = 0 to [Mg/Fe] = 1). However, the gravity drops to  $\log g$ = 0.5 if we set 
[Mg/Fe] = 1.8. Thus, the effect is indeed very large if Mg is as abundant as oxygen. However, as we 
have already stated, the Mg abundance derived from the 4571 \AA\ line ([Mg/Fe]=1.8) is most 
probably overestimated and therefore we do not find it necessary to revise the stellar parameters obtained 
for [Mg/Fe] = 0 or [Mg/Fe] = 1. The final parameters for CS\,29498--043 are assumed $T_{\rm eff}$ = 4400 K, 
$\log g$ = 1.5 and [Fe/H] = $-3.5$ when [Mg/Fe] = 1.0.

\section{Discussion}

Our results may have an interesting impact on the physics of the supernova progenitor which 
gave birth to these ultra-metal-poor stars. Should we use the abundance from the triplet 
or the forbidden line to set constrains on the supernova models ? The situation with our targets
clearly demonstrates that non-LTE effects on the near-IR oxygen triplet are not 
responsible for this conflict, and that therefore this abundance indicator is as reliable 
(or unreliable) as the forbidden line. 

Both CS\,22949--037 and CS29498--043 are distinguished from other carbon-rich metal-poor
stars with large excesses of neutron capture s-process elements. The overabundance
of s-elements is usually attributed to nucleosynthesis in the thermally pulsing AGB stars.
These ultra-metal-poor giants have moderately high abundances of Mg, Si and Al, while
C and N are unusually overabundant. In this paper we find that both stars are also rich
in oxygen. It appears that the class of objects with high [O/Fe] is not limited by CS\,22949--037
and CS29498--043.  Aoki et al. (2002) proposed the existence of a new 
class of very metal-poor stars which originate from supernova in which most of the matter
were absorbed by the iron core. The most metal-poor star in the Galaxy, HE0107--5240, 
also belongs to this class of objects and has a large [O/Fe]=2.4 derived from the UV OH lines 
(Bessell et al. 2004). The abundance pattern in this star is consistent with a model 
(Umeda \& Nomoto 2003) in which the supernova undergoes some mixing followed by a fallback
into a massive black hole. However, the oxygen abundance of HE0107--5240 is smaller than
the prediction of Umeda \& Nomoto (2003) by as much as 1 dex. The prototype of this type of 
supernova is SN1997D, which 
was very underluminous because of the small amount (2$\times10^{-3}$M$_{\sun}$) of $^{56}$Ni
ejected during explosion. Four new faint supernova have been reported recently 
(Pastorello et al. 2003). While such supernovae are not observed frequently, their real number  
is expected to be much higher because of the faintness of the supernovae.
These faint supernovae are expected to be more frequent in the early Galaxy (Umeda \& Nomoto 2003),
while their ejecta are characterized by very high [O/Fe] ratios. In fact, the monotonically rising 
trend of [O/Fe] can possibly be explained if we assume that the iron yield decreases with
stellar mass (i.e. most massive stars form massive black holes). It is clear that the 
formation rate of massive black holes in the early Galaxy may affect the observed [O/Fe] trend. 
A massive black hole in the low mass X-ray binary system Nova Sco 1994 with [O/Fe] = 1.0
(Israelian et al. 1999) may serve as a prototype for such ``failed'' supernovae, where almost all the 
Fe has been accreted by the black hole. Apparently CS22949-037 is not an ``exceptional" star as
noted by Depagne et al.\,(2002). There are another four high [O/Fe] stars: CS29498-043, 
HE0107-5240, CS29497-030 (Sivarani et al.\,2003) and LP625-44 (Aoki et al. 2002b). Abundances
in stars such as LP625-44 and CS29497-030 are assumed to result from mass transfer from an AGB
star across a binary on to the observed companion star. However, it is not clear whether this explains
the [O/Fe] excess in these s-process-enhanced stars. Other s-process-rich stars such as LP706-7 
(Norris et al. 1997) do not show any radial velocity variations. There is much work to be done
before we will be able to understand why some s-process stars are also oxygen rich. Displaying these stars 
on the [O/Fe] versus [Fe/H] diagram with a representative sample of halo dwarfs and giants 
from the literature (e.g. Nissen et al.\,2002, Cayrel et al.\,2003, Israelian et al.\,2001), 
we can see the general trend of oxygen in the galaxy and the relative position 
of these ``extreme" stars with respect measurements of other ``normal" stars (Fig. 15). This
diagram gives a broader perspective on the evolution of oxygen in the Galaxy.
We propose the existence of an upper envelope of the [O/Fe] ratio represented by the dashed line
which suggests a monotonically increasing trend toward lower metallicities (Fig. 15). 
Very high [O/Fe] ratios are possibly   indicating  that most of the Fe nuclei synthesized in 
the inner core are actually held by a massive compact object, i.e a black hole. 
Smaller ratios possibly indicate that a significant fraction of the Fe nuclei is incorporated 
into the supernova ejecta, therefore  a smaller mass cut is required and the likely formation of 
less massive compact objects, i.e neutron stars. Assuming that below metallicity $-$3, we are 
seeing direct yields from the first generation of supernovae,  the range in [O/Fe] encompassed  
by the dashed lines in our plot may be populated by stars contaminated by supernovae that led
to the formation of compact objects with different masses.

The fact that the atmospheres of these stars do not contain large amounts of Ca, Ti, Si and Mg 
supports the idea of massive black holes left from the supernova explosions of
the first stars. It has been suggested that the bulk of light r-nuclei (with $A$ $<$ 130) appear to 
have different sources from those for heavy r-nuclei (Wasserburg, Busso \& Gallino 1996). 
The meteoritic data require at least two distinct types of SN r-process events: the high-frequency 
events, $H$, producing heavy nuclei with $A$ $>$ 130, including $^{182}$Hf, and the
low-frequency $L$ events producing light nuclei with $A$ $>$ 130, including $^{129}$I. The
r-process production in the SN environments associated with the $H$ and $L$ events
has been discussed in some detail by Wasserburg \& Qian (2000). The abundance analysis of 
CS\,22949--037 and CS\,29498--043 extended to heavy neutron capture elements 
may directly test the speculation by Wasserburg and Qian (2000) that $H$ events are associated with 
supernovae producing black holes, whereas  $L$ events are associated with supernovae producing 
neutron stars. According to this model, the parent supernovae of our targets come from the
$H$-events.

The large disagreements found from different abundance indicators of Mg and O reveal that
the atmospheric models used in this study are not reliable. The conflict is so severe 
that we cannot question the quality of the data and/or
the model atoms used in our non-LTE. The problem with the oxygen and magnesium abundances may have the same
roots as that discussed by Dalle Ore (1993). This author found large discrepancies 
among the temperatures obtained from the excitation and ionization equilibrium of several cool giants.
In fact, the systematic disagreement between different temperature scales found from the 
continuum energy distribution, H$\alpha$ and Fe lines is the best indication that the
models employed in these studies are to some extent unreliable. 

Our analysis suggests that the gravities of very metal-poor giants derived from the LTE 
Fe analysis are strongly underestimated because non-LTE effects are neglected. The oxygen abundances 
in CS\,22949--037 and CS\,29498--043 derived from the triplet and the forbidden line differ 
by a large factor. It is interesting that the oxygen 
forbidden line at 6300 \AA\ and the near-IR triplet are formed in the same layers deep in the 
atmosphere (Fig. 9). This suggests that one needs a different atmospheric structure in order to 
achieve consistency for O and Mg. It is clear that the standard 1D models of Kurucz (1992) 
are unreliable for ultra-metal-poor giants. As a final check of these results, we used models
without convective overshooting  (Castelli, Gratton \& Kurucz 1997) and found that
they do not resolve the discrepancy either.

\section{Conclusions}

Observations with Keck I/HIRES have revealed strong lines of the oxygen near-IR triplet
in the spectra of the ultra-metal-poor giants CS\,22949--037 and CS\,29498--043.
The forbidden line of oxygen with EW = 60$\pm$10m\AA\ was observed with TNG/SARG.
A detailed non-LTE analysis of Fe has been carried out and a new set of the atmospheric
parameters have been obtained. Our analysis suggests that the gravities of metal-poor giants 
derived from the LTE Fe analysis are strongly underestimated because of the neglect of non-LTE effects.

The oxygen abundance in CS\,22949--037 and CS29498--043 derived from the triplet and the forbidden line 
differ by 1.18 and 0.53 dex, respectively. This disagreement cannot be explained by a non-LTE effects, 
quality of the data and/or uncertainties in stellar parameters. Other mechanisms must be invoked in order 
to explain this puzzle. A similar discrepancy was found for Mg when comparing the abundances obtained from 
the resonance line 4571 \AA\ and several strong subordinate lines. Based on the present analysis 
we propose that the Kurucz (1992) models are not reliable for these ultra-metal-poor giants.

\section{Acknowledgments}

The data presented here were obtained with the W. M. Keck Observatory,
which is operated as a scientific partnership among the California
Institute of Technology, the University of California, and the
National Aeronautics and Space Administration. The Observatory
was made possible by the generous financial support of the
W. M. Keck foundation. We are grateful to Wako Aoki and Martin Asplund for several helpful 
discussions and Piercarlo Bonifacio for providing interpolated models of Castelli et al. 
without overshooting. We thank the anonymous referee for useful suggestions and comments. 
N. S. would like to thank R.~Kostik for several discussions and Irina Vasiljeva for a 
help with computations. This research was partially supported by the Spanish DGES under project 
AYA2001-1657 and by INTAS grant 00-00084.

{}

\end{document}